\documentclass[aps,apl,preprint,superscriptaddress]{revtex4}
\usepackage{graphicx}
\usepackage{epstopdf}
\begin{document}

\title{Real-space characterization of cavity-coupled waveguide systems\\in hypersonic phononic crystals}
%\title{Demonstration of wavelength-scale and multi-modal cavities\\in hypersonic crystal slabs and their excitation via a single-mode waveguide}
%\title{Systematic study on multi-modal hypersonic-crystal cavities \\and cavity-waveguide systems}

\author{D. Hatanaka}

\email{daiki.hatanaka.hz@hco.ntt.co.jp}

\author{H. Yamaguchi}

\affiliation{NTT Basic Research Laboratories, NTT Corporation, Atsugi-shi, Kanagawa 243-0198, Japan}

\begin{abstract}
A phononic crystal formed in a suspended membrane provides full confinement of hypersonic waves and thus realizes a range of chip-scale manipulations. In this letter, we  demonstrate the mode-resolved real-space characterization of the mechanical vibration properties in cavities and waveguide systems. Multiple resonant modes are independently characterized in various designed cavities, and wavelength-scale high-$Q$ resonances up to $Q=$ 4200 under atmospheric conditions are confirmed. This also reveals that the waveguide allows us to resolve single-mode wave transmission and thereby drive evanescently-coupled cavities. The methods offer a significant tool with which to build compact and low-power microwave phononic circuitry for signal processing and hybrid quantum system applications.  
\end{abstract} 

\maketitle

\hspace*{0.5em}The on-chip control of acoustic phonons at microwave frequencies, namely \textit{hypersound}, has been essential for a wide range of applications such as sensing for bio-chemical detection \cite{saw_bio}, analog signal processing \cite{morgan_saw} and wireless communications \cite{hashimoto_saw}. Moreover, this has also realized an alternative way to drive another degree of freedom such as photons \cite{fuhrmann_saw, tadesse_saw, balram_optomecha, painter_nphoton, gustafsson_saw, chu_qae, manenti_qae, noguchi_saw, schoelkopf_saw, cleland_saw, sletten_saw, safavi_pnc1, hypersonic_painter}, electrons \cite{naber_saw, gustafsson_saw2, fujisawa_saw1, fujisawa_saw2} and spins \cite{weiler_adfmr, weiler_spinpump, sanada_saw, thevenard_adfmr1, labanowski_adfmr, kobayashi_src, golter_saw_nv, whiteley_nvspin}, thus giving rise to development of novel hybrid architectures. In the above systems, hypersonic vibrations can be exploited to efficiently actuate and modulate these energy particles, and to inter-connect information between different sub-systems, which offers new opportunities in emerging fields such as quantum-acoustics \cite{gustafsson_saw, chu_qae, manenti_qae, noguchi_saw, schoelkopf_saw, cleland_saw, sletten_saw, safavi_pnc1} and spin-mechanics \cite{weiler_adfmr, weiler_spinpump, sanada_saw, thevenard_adfmr1, labanowski_adfmr, kobayashi_src, golter_saw_nv, whiteley_nvspin}. It is becoming increasingly important to develop the ability to manipulate such ultrahigh-frequency acoustic-phonons and thus develop phononics technology.\\
\hspace*{1.0em}A phononic crystal (PnC) is an artificial material consisting of periodic elastic composites \cite{maldovan_nature, narayanamurti_pnc, martinez_pnc, benchabane_pnc1,hatanaka_pnc}. By using this engineered structure, the dispersion relation and bandgap of acoustic waves can be tailored and formed, which enables acoustic propagation to be controlled. For instance, a local modulation in the periodic geometry creates a cavity \cite{mohammadi_pnc1} and a waveguide \cite{otsuka_pnc, benchabane_pnc2, adibi_pncslab, baboly_pncslab2} sustained by the bandgap, which spatially traps and guides acoustic waves, respectively. This PnC concept has been introduced into various micromechanical systems at microwave frequencies, including surface acoustic wave (SAW) devices \cite{tang_sawpnc, loncar_sawpnc}, suspended nanobeams \cite{balram_optomecha, painter_nphoton, safavi_pnc1, hypersonic_painter, painter_pncQ} and membranes \cite{adibi_pncslab, baboly_pncslab2, baboly_pncslab1}. In terms of the scalability of device integration, a suspended PnC membrane is a suitable platform because it can realize both a high quality ($Q$) factor and a small mode volume ($V_{\rm m}$), leading to a low loss and a small device footprint. This also allows us to incorporate various components into a two-dimensional membrane, thus enabling intricate phonon manipulations. However, the key building blocks such as cavities, waveguides and their evanescently-coupled structures based on the PnC membrane have yet to be fully investigated. In particular, the lack of information on real-space vibration structures of various device geometries prevents mode structure, volume and wavenumber from being precisely determined. This difficulty limits the systematic characterization of this platform and thus its availability.\\
\hspace*{1.0em}In this letter, we fabricate and investigate PnC cavities and waveguides built in suspended GaAs membranes. Measuring the spectral responses enables us to find multiple hypersonic resonances in the 0.50-0.58 GHz range and then, their frequencies and $Q$-factors. The real-space visualization of their vibrations using an optical interferometric technique identifies the modal shapes that allow the mode volumes and wavenumbers to be extracted by comparison with a calculation using the finite-element method (FEM). As a result, we realize a PnC cavity with $Q=$ 4200 and $V_{\rm m}=$ 2.4 $\mu$m$^{3}$ = 0.18$\lambda^{2}t$ at 0.5435 GHz in the atmosphere, where $\lambda$ and $t$ are the acoustic wavelength and membrane thickness, respectively. The high $Q$ and wavelength-scale $V_{\rm m}$, respectively, are comparable to and smaller than those of conventional PnC cavities at room temperature \cite{balram_optomecha,painter_nphoton,mohammadi_pnc1,baboly_pncslab1}. Furthermore, we fabricate and investigate cavity-waveguide systems where a cavity is evanescently coupled to a waveguide and show that the over- and under-coupling conditions are changed by the spatial separation between them. Our results provide useful information for designing a cavity and a waveguide and for constructing compact and low-power phononic circuitry for opto-mechanics and spin-mechanics applications.\\
\hspace*{1.0em}The device has a GaAs (1.0 $\mu$m) / Al$_{0.7}$Ga$_{0.3}$As (3.0 $\mu$m) heterostructure as shown in Fig. 1(a). Periodic arrays of air-holes are patterned in the GaAs layer, which is suspended by sacrificially etching the Al$_{0.7}$Ga$_{0.3}$As layer through the holes with diluted hydrofluoric acid (see the left insets in Fig. 1(b)). A triangular lattice of snowflake-shaped holes (right inset in Fig. 1(b)) is used for the PnC geometry (middle inset) \cite{safavi_snowflake} and this gives rise to a complete phononic bandgap in the 0.5-0.6 GHz and 0.7-0.8 GHz regions by adopting the micrometer-scale periodicity shown in Fig. 1(c). Hypersonic waves are piezoelectrically excited from one of the inter-digit transducers (IDTs) found on both sides of the device by applying microwave signals to it. The acoustic waves travel on the surface of the GaAs layer as Rayleigh waves and are then transformed into Lamb waves in the PnC membrane (see section 1 in Supplementary Material). All measurements were performed using an optical interferometer in the atmosphere. The approach is used for real-space characterization of the devices by scanning of a laser over them laterally.\\
\hspace*{1.0em}First, we characterize resonance modes in a PnC cavity, which is formed by removing holes from a periodic lattice, as the simplest and most basic device structure. Figure 2(a) shows an L3 cavity from which three in-line holes are missing. Such a line-defect cavity has the potential to enable hypersonic waves at frequencies within the bandgap to be confined in the defect. To confirm the possibility, the spectral response of the L3 cavity is measured under piezoelectric excitation from one of the IDTs. The electromagnetic wave reflection from the IDT ($S_{11}$) exhibits distinct absorption around 0.53 GHz as shown in the top panel of Fig. 2(b), indicating that the injected energy is converted to SAW (Rayleigh wave) by the piezoelectric effect. The IDT operation bandwidth ($f_{\rm SAW}$) can be estimated from $f_{\rm SAW}=v_{\rm GaAs}/p$ = 0.521 GHz, where $v_{\rm GaAs}$ and $p$ are the SAW velocity of GaAs (2867 m/s \cite{gaas_vsaw}) and a periodic pitch of IDT (5.5 $\mu$m), respectively. The slight deviation from the experimental value might be due to the different SAW velocity of the Al$_{0.7}$Ga$_{0.3}$As underlayer \cite{gaas_vsaw}. The cavity dynamics are evaluated from an $S_{21}$ transmission in the bottom panel of Fig. 2(b), which reveals a defined sharp peak around 0.529 GHz. By fitting a Lorenz curve to the peak as shown in the inset, the $Q$ factor can be estimated at $Q$ = 3700. Real-space mapping is performed by scanning the laser spot position over the cavity at 0.5293 GHz, and the obtained amplitude and phase components are plotted as shown in the top and bottom panels of Fig. 2(c), respectively. Indeed, the vibration is well localized in the defect area and shows a certain mode shape. The resultant amplitude and phase distributions of this mode can be predicted from FEM using COMSOL Multiphysics as shown in the insets of Fig. 2(c). The simulations indicate that this mode is induced by an asymmetric Lamb wave. This also allows the effective mode volume to be calculated by integrating the vibration energy over the entire membrane as $V_{\rm m}$ = $\int d^{2}r$$(\frac{|q(r)|}{{\rm max}(|q(r)|)})^{2}t$ = 8.1 $\mu$m$^{3}$ = 0.63$\lambda^{2}t$, where $q(r)$ is the mode profile displacement. Thus, the results indicate that the line-defect cavity formed in the snowflake PnC lattice realizes the wavelength-scale lateral confinement of hypersonic vibrations and a $Q$ factor as high as that reported in conventional hypersonic cavities at room temperature \cite{tang_sawpnc,loncar_sawpnc,balram_optomecha,painter_nphoton,baboly_pncslab1}.\\
\hspace*{1.0em}In this way, the spectral response of five types of L$x$ cavity ($x$ = 1 $\sim$ 5) is investigated over a broad frequency span of 0.5-0.6 GHz. Finally, we observe multiple hypersonic resonances in each cavity and also the corresponding frequencies and $Q$s. The obtained $Q$s in the L2 cavity are plotted with respect to the resonant frequencies as shown in Fig. 3(a). A total of five different modes can be found thanks to the real-space mapping measurements and they are labeled as M1 to M5 denoted by circles of different colors. By repeating the same measurements with several samples, we confirm that each mode shows a similar resonant frequency. To support the findings, FEM simulations are conducted. The simulation results reveal that the elastic anisotropy of GaAs crystal plays an essential role in explaining the mode frequencies. Figure 3(a) shows the results using both isotropic and anisotropic elastic parameters as denoted by dashed and solid lines, respectively \cite{adachi_gaas}. Note that calculated mode frequencies can predict the experimental ones when considering the elastic anisotropy, whereas the results obtained using the isotropic parameters differ significantly from the experimental results. Similar results are also obtained in other L1, L3 and L4 cavities (see section 2 in Supplementary Material).\\
\hspace*{1.0em}The $Q$ factor describing energy loss is an essential parameter for evaluating cavity characteristics. All the $Q$s experimentally obtained from the L1-L5 cavities are classified as different modes (M1-M5) based on real-space mapping results as shown in Fig. 3(b). Although there is a finite variation in the $Q$s at each mode, the M1 mode is likely to show a higher $Q$ factors (average $Q_{\rm cav}$ = 2080) than the others ($Q_{\rm cav}$ = 1286-1447). Next, we focus on the M1 mode in an L1 cavity and investigate the $Q$ factor at various numbers of acoustic shield periods ($N$) as shown in Fig. 3(c). The $Q$ values increase monotonically with increasing $N$ and in particular, the highest $Q_{\rm cav}$ = 4200 and smallest $V_{\rm m}=$ 2.4 $\mu$m$^{3}$ = 0.18$\lambda^{2}t$ in this work are found in $N$ = 7 denoted by a filled star and its spectral response is shown in Fig. 3(d). However, the measured $Q$ increment with $N$ is much less than that predicted by FEM calculations (see section 3 in Supplementary Material). Therefore, there are another major dissipation sources. Material loss such as thermoelastic damping and a surface two-level system (TLS) possibly induce dissipation \cite{cleland_nems}, whereas air-damping is not significant because we confirm that the $Q$ values do not change in a vacuum. According to a previous report on GaAs optomechanical systems operating at 0.26 GHz by Hamoumi et al. \cite{hamoumi_gaas}, the major dissipation is caused by a GaAs TLS at room temperature. It ranges from 50-100 kHz and is the same order of magnitude as that of our cavities. Thus, internal dissipation such as TLS can also be dominant in our devices, and surface treatment during or after fabrication could help to increase the $Q$ factors.\\
\hspace*{1.0em}Secondly, an acoustic waveguide is realized by creating a line defect in the periodic lattice, where hypersonic waves are guided while being sustained by the bandgap. Figure 4(a) shows a PnC waveguide formed by simply removing one row of holes from the triangular lattice. An incident Lamb wave excited from an IDT can be confined in the line defect and spatially guided as shown in the bottom panels of Fig. 4(a). This waveguiding mode has one vibration antinode in the center of the waveguide and it is calculated as mode profile A in Fig. 4(b). The visualized spatial evolutions of the amplitude and phase provide wavenumber information about the waves. In this way, the dispersion relation of acoustic propagations in the waveguide can be constructed by collecting these data at various frequencies of 0.50-0.58 GHz, and this is plotted as the solid red circles in Fig. 4(c). Also shown is the dispersion relation calculated by FEM, where phononic bands for Lamb waves (out-of-plane displacement) and Love waves (in-plane displacement) are depicted in color (light blue, green and red) and gray, respectively. Because of the limitation of our optical detection method, we can only detect out-of-plane Lamb waves and thus, single-mode waveguiding in branch A can be resolved in the above frequency range. Additional branches B and C are available as depicted by light green and red, and can support waveguiding modes containing two and three antinodes, which can be seen in Fig. 4(b), respectively. In particular, branch A interacts with branch C showing anti-crossing around 0.61 GHz, but not with branch B due to the parity difference between odd and even modal shapes.\\
\hspace*{1.0em}The use of the waveguide enables us to spatially address distant cavities. Such a cavity-waveguide coupled geometry is fundamental to the construction of integrated phononic circuits as shown in the left inset in Fig. 5(a). Here, we use L2 cavity-waveguide systems with various cavity-waveguide separations $N_{\rm c}$ = 1, 2 and 3 between these two components as shown in the left, middle and right panels of Fig. 5(a), respectively, and measure the spectral response of the cavities as shown in the left panels of Fig. 5(b). All the cavities exhibit a defined resonance peak around 0.534 GHz, which is indeed sustained by the M1 mode as confirmed by the right panels in Fig. 5(b). The average $Q$ factors at these three configurations are examined, which reveals that the loaded $Q$ factor of the cavity is $Q_{\rm load}$ = 600 in $N_{\rm c}$ = 1 and is lower than the others in $N_{\rm c}$ = 2 ($Q_{\rm load}=$ 1474) and $N_{\rm c}$ = 3 ($Q_{\rm load}=$ 1454) as shown in Fig. 5(c). On the other hand, it is invariant between $N_{\rm c}$ = 2 and $N_{\rm c}$ = 3. Briefly, the dissipation of a cavity in such a cavity-waveguide system is given by,
\begin{equation}
Q_{\rm load}^{-1}=Q_{\rm cav}^{-1}+Q_{\rm wg}^{-1},
\end{equation}
where $Q_{\rm wg}$ is the quality factor for the dissipation through the waveguide. From the experimental result in Fig. 3(b) indicating the intrinsic cavity $Q$-factor in M1 mode ($Q_{\rm cav}=2080$), $Q_{\rm wg}$ can be extracted to 843, 5059 and 4831 in $N_{\rm c}$ = 1, 2 and 3 configurations, respectively. Therefore, the cavity-waveguide system with $N_{\rm c}$ = 1 is in an over-coupling regime, namely $Q_{\rm wg}^{-1}>Q_{\rm cav}^{-1}$, where energy dissipation through the waveguide is dominant and in other words, it enables a cavity to be efficiently driven. On the other hand, the other configurations with $N_{\rm c}$ = 2 and 3 realize an under-coupling regime, $Q_{\rm wg}^{-1}<Q_{\rm cav}^{-1}$, where the intrinsic cavity dissipation is dominant, and thus a cavity can be excited while retaining the intrinsic $Q_{\rm cav}$. Thus, the coupling condition between a cavity and a waveguide can be altered with spatial separation, and the transition from over- to under-coupling can be found between $N_{\rm c}$ = 1 and 2.\\
\hspace*{1.0em}This cavity-waveguide structure is a building block for the construction of phononic circuits. For instance, an acoustic microwave filter composed of cavities sandwiched by two waveguides is demonstrated, as shown in Fig. 6(a). Hypersonic waves at 0.561 GHz excited from an IDT are injected into the input waveguide (bottom) and then propagate to the output waveguide (top) through the cavities, which can be visualized as shown in Fig. 6(b). The resultant waves in the output waveguide are filtered by the cavity resonance, the bandwidth of which is $\sim$1 MHz as determined by $Q$ factor. This bandpass region is narrower than that of the input waveguide $\sim$5 MHz resulting from the spectral response of the IDT. In conventional microwave filters based on SAW devices, the bandwidth is given by the number of IDT electrodes. Namely, a narrower passband can be realized with a larger number of electrodes on a SAW filter. Briefly, 500 electrode finger-pairs are needed to fabricate a 1-MHz bandpass filter that has an IDT with a length of several-millimeters. On the other hand, a filter based on a PnC membrane realizes a high $Q$ and a low $V_{\rm m}$ and so it can be band-narrowing and wavelength-scale compact. By optimizing IDTs with high peizoelectric materials such as LiNbO$_{3}$ and PZT, the finger-pair number can be reduced, and thus, the entire device size can be made much smaller with a length of several tens of micrometers. Additionally, the bandwidth and spectral position are adjustable by engineering the cavity geometry. For instance, a broad passband can also be realized by forming a linear cavity array. Moreover, the propagation loss will be suppressed by optimizing the coupling geometry and modal shapes of the cavity and the waveguide. This hypersonic PnC holds promise as a platform on which to develop functional and compact phonon circuits.\\
\hspace*{1.0em}In conclusion, hypersonic PnC devices in a suspended GaAs membrane are demonstrated and their mechanical vibration properties are investigated with a real-space mapping technique. Experiments on real-space characterization reveal that multiple hypersonic resonances are available in various line-defect cavities, and in particular, a high quality factor $Q=4200$ and a small mode volume $V_{\rm m}=0.18\lambda^{2}t$ are realized in an L1 cavity. Additionally, the dispersion relation of a waveguide is constructed by acquiring the amplitude and phase of spatial information, clarifying the possibility of realizing single-mode waveguide systems. Moreover, this waveguide allows us to drive a cavity remotely, and the driving efficiency can be changed by adjusting the coupling geometry. The PnC architecture enables the realization of compact and functional phononic circuitry, thus increasing the availability of acoustic phonons in conventional signal processing applications as well as opto-mechanics and spin-mechanics.\\
\\
\hspace*{0.5cm}We thank Drs. T. Tamamura, M. Ono, S. Sasaki and Y. Harada for their support and helpful comments on device fabrication. We are also grateful to Drs. R. Ohta and M. Asano for fruitful discussions on FEM calculation. This work was supported by a MEXT Grant-in-Aid for Scientific Research on Innovative Areas “Science of hybrid quantum systems” (Grants No. JP15H05869 and JP15K21727). We acknowledge the stimulating discussion at the meeting of the Cooperative Research Project of the Research Institute of Electrical Communication, Tohoku University.

%\bibliography{ref_ghz-pnc}

\newpage

\begin{figure}[b]
	\begin{center}
		\vspace{-0.2cm}\hspace{-0.0cm}
		\includegraphics[scale=0.45]{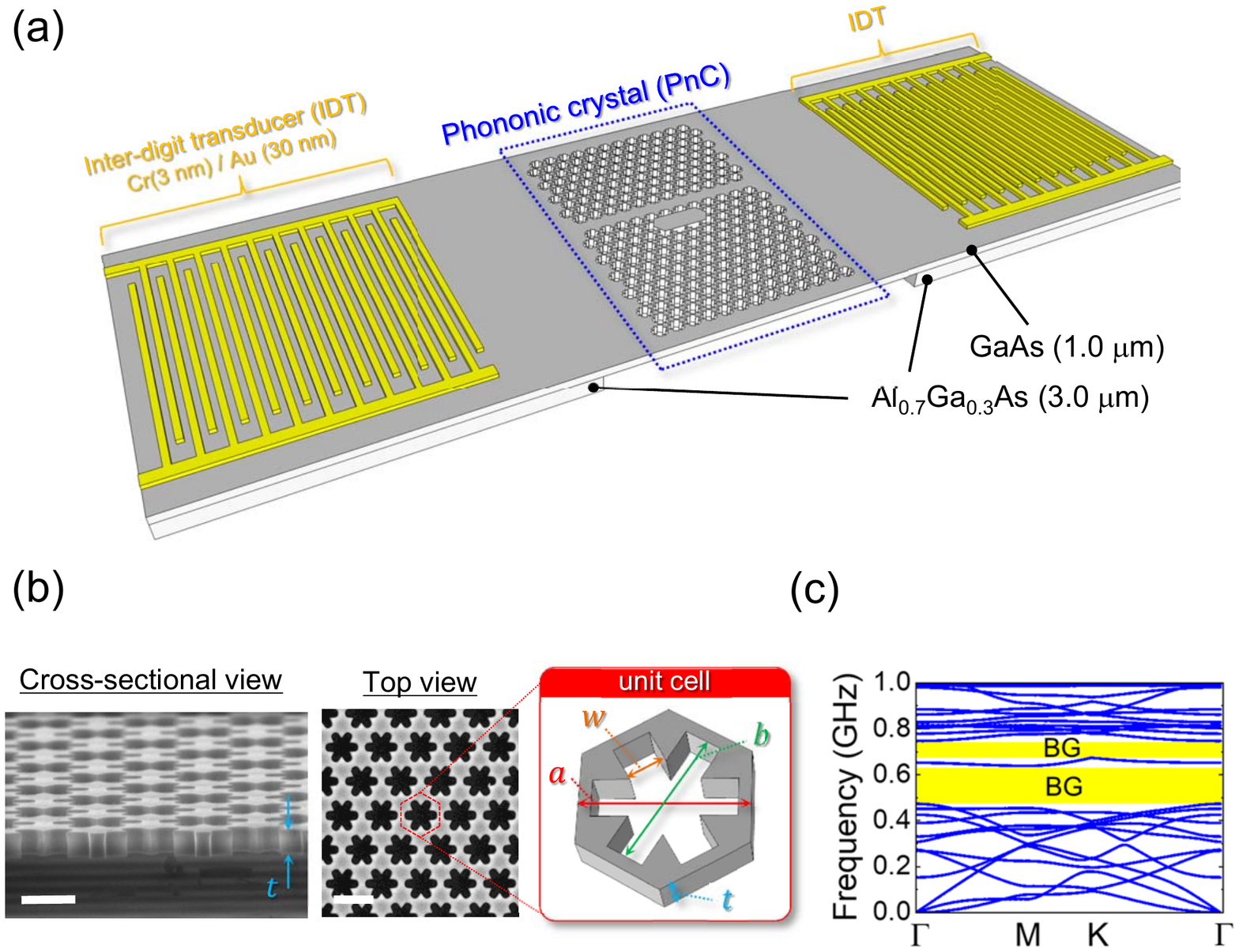}
		\vspace{-6.5cm}
		\caption{\textbf{(a)} A schematic of an electromechanical PnC device, in which a two-dimensional phononic shield is formed in a 1.0 $\mu$m-thick GaAs membrane by forming a triangular lattice with periodic air holes. IDTs with Cr (3.0 nm) / Au (30 nm) are introduced on both bulk joint parts that enable hypersonic waves to be piezoelectrically excited with the application of electromagnetic waves. Al$_{0.7}$Ga$_{0.3}$As (3 $\mu$m) is a sacrificial layer and selectively etched to form the suspended GaAs membrane. \textbf{(b)} An SEM image of the cross-section and top of the PnC lattice is shown as the left and middle insets, where the scale bars are 2 $\mu$m and 4 $\mu$m, respectively. A unit cell containing a snowflake-shaped air hole is shown in the right inset, where $a$ = 4.0 $\mu$m, $b$ = 3.4 $\mu$m, $w$ = 1.0 $\mu$m and $t$ = 1.0 $\mu$m are the lattice constant, hole length, hole width and membrane thickness, respectively. \textbf{(c)} Dispersion relation of the PnC lattice calculated by FEM exhibits complete bandgaps (BG) at 0.5-0.6 and 0.7-0.8 GHz, where hypersonic waves cannot propagate.}
		\label{fig 1}
		\vspace{-1cm}
	\end{center}
\end{figure}

\newpage

\begin{figure}[t]
	\begin{center}
		\vspace{-0.2cm}\hspace{1.0cm}
		\includegraphics[scale=0.51]{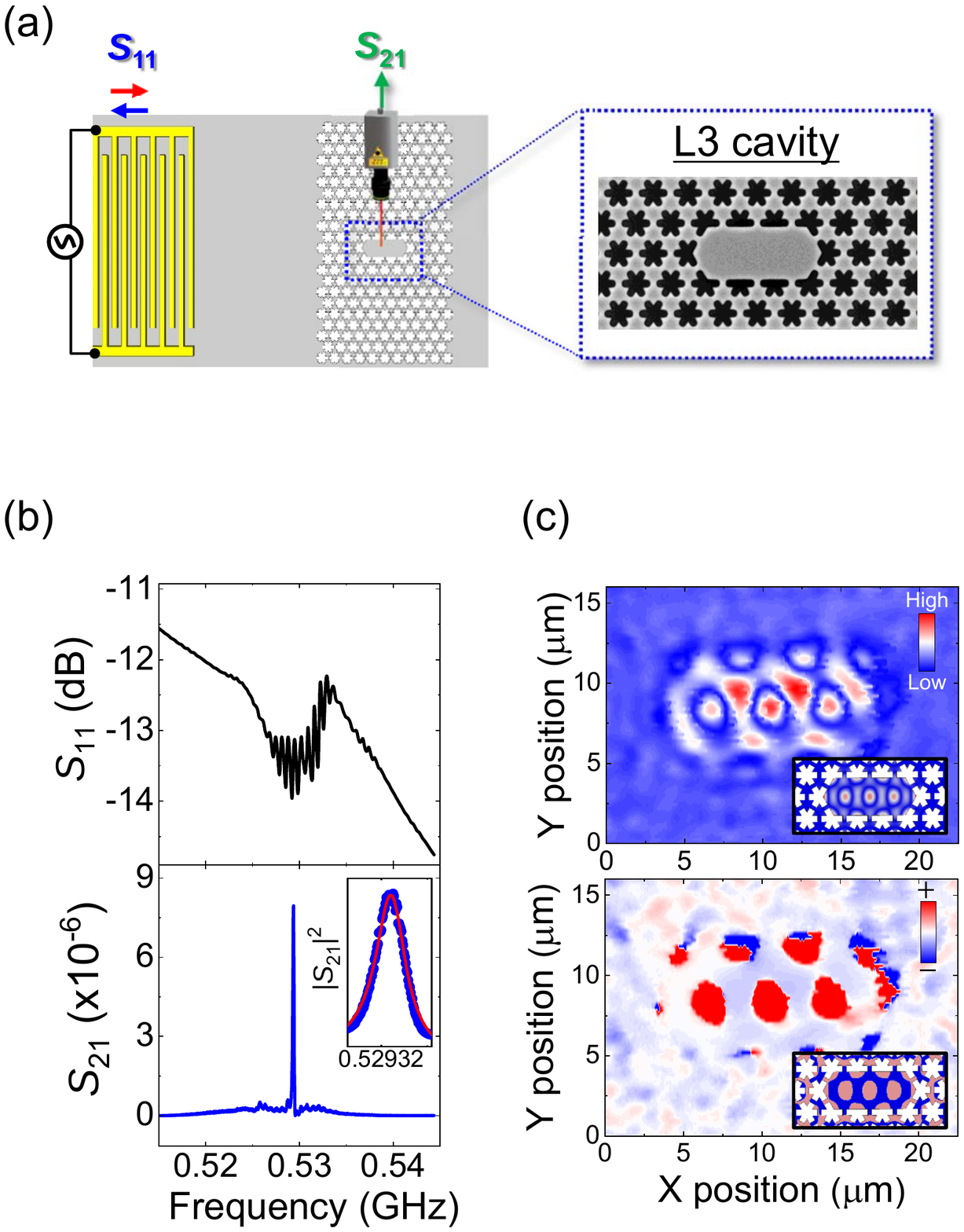}
		\vspace{-4.0cm}
		\caption{\textbf{(a)} The measurement setup for a line-defect (L3) cavity. The application of electromagnetic waves to an IDT electrode excites SAWs (Rayleigh waves) through the piezoelectric effect. The acoustic waves are sent into the cavity, which is surrounded by a PnC lattice. The resultant vibration spectrum is probed using an optical interferometer under atmospheric conditions. \textbf{(b)} Spectral responses of the electrical reflection from the IDT ($S_{11}$, top panel) and the optically measured transmission ($S_{21}$, bottom panel) of the device are shown. The inset shows the zoomed-in spectral response of $|S_{21}|^{2}$ and the red solid line shows a fitting curve. \textbf{(c)} Real-space mapping of the amplitude and phase of the vibration at 0.5293 GHz is shown in the top and bottom panels, respectively, and the corresponding mode profiles simulated by FEM are also shown in the insets.}
		\label{fig 2}
		\vspace{-0.7cm}
	\end{center}
\end{figure}

\newpage

\begin{figure}[t]
	\begin{center}
		\vspace{-0.2cm}\hspace{-2.9cm}
		\includegraphics[scale=0.48]{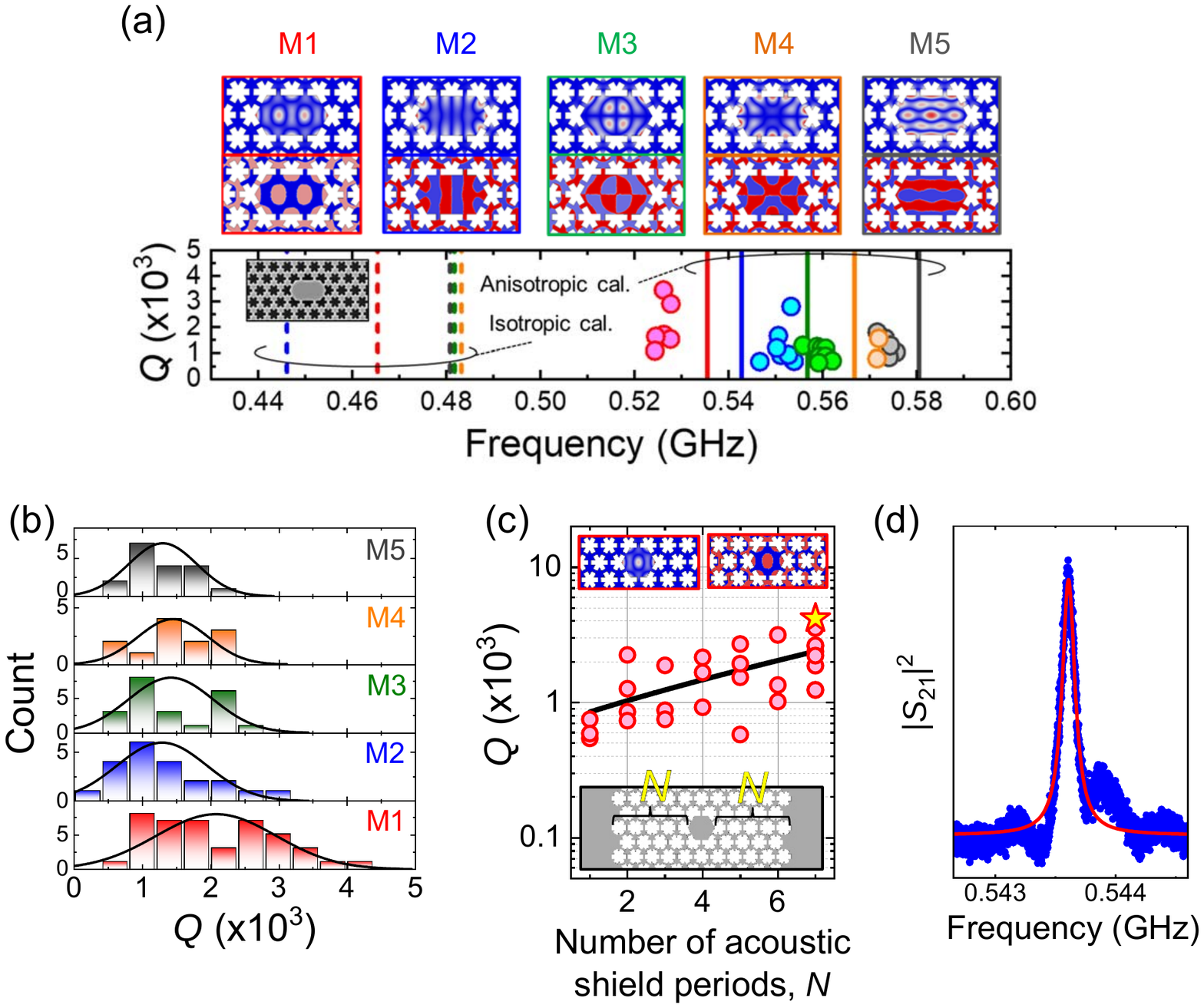}
		\vspace{-6.5cm}
		\caption{\textbf{(a)} Resonance frequencies and $Q$ factors of multimodal L2 cavities (M1, M2, M3, M4 and M5) with $N=$ 5-7 are plotted as red, blue, green, orange and gray circles, respectively. The resonance frequencies are numerically calculated by including anisotropic and isotropic elastic constants as shown with solid and dashed lines, respectively. The top panels show the amplitude (top) and phase (bottom) distributions of the mode profiles of M1-M5 by FEM. \textbf{(b)} Histogram of $Q$ factors in M1-M5 and the fitting curves using a Gaussian distribution function as shown from the bottom to top panels, respectively. The experimental data are taken from L1-L4 cavities with $N=$ 5-7. \textbf{(c)} Number of acoustic shield period ($N$) dependence of $Q$ factors of the M1 mode in an L1 cavity . The solid black line is an exponential fitting. The mode profiles of amplitude and phase are depicted in the left and right panels of the top inset, respectively. A schematic of the device is shown in the bottom inset where $N$ is defined. \textbf{(d)} The spectral response of the L1 cavity with $N=$ 7 (blue) as denoted by a star in (c) and a ftting curve (red), showing the highest $Q=$ 4200.}
		\label{fig 3}
		\vspace{-0.5cm}
	\end{center}
\end{figure}

\newpage

\begin{figure}[b]
	\begin{center}
		\vspace{-0.2cm}\hspace{1.0cm}
		\includegraphics[scale=0.44]{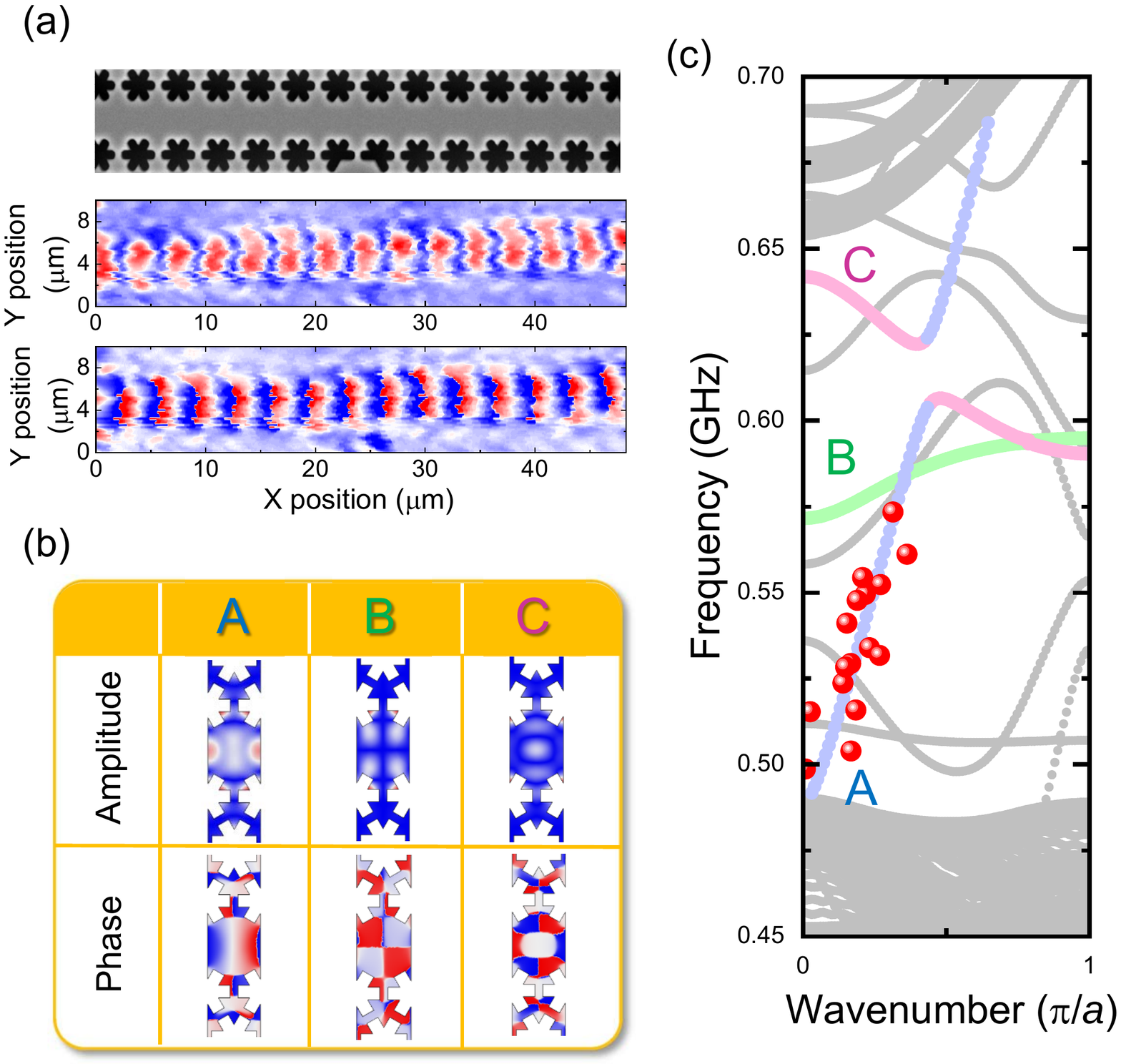}
		\vspace{-4.5cm}
		\caption{\textbf{(a)} An SEM image of a line-defect waveguide (top), and its amplitude (middle) and phase (bottom) spatial distributions at 0.573 GHz. \textbf{(b)} Mode profiles of the guiding modes A, B, and C simulated by FEM. \textbf{(c)} Dispersion relation of a line-defect waveguide in a snowflake-based PnC. The FEM calculation shows multiple guiding modes in the bandgap of 0.50-0.65 GHz, where several asymmetric Lamb modes (light blue, green and red lines) and Love modes (gray line) exist simultaneously. The phononic branches of the Lamb modes are labeled A, B, and C, which correspond to the mode profiles shown in (b). The dispersion relation is also experimentally estimated from the real-space mapping results as denoted by solid red circles.}
		\label{fig 4}
		\vspace{0.0cm}
	\end{center}
\end{figure}

\newpage

\begin{figure}[t]
	\begin{center}
		\vspace{-0.5cm}\hspace{2.0cm}
		\includegraphics[scale=0.45]{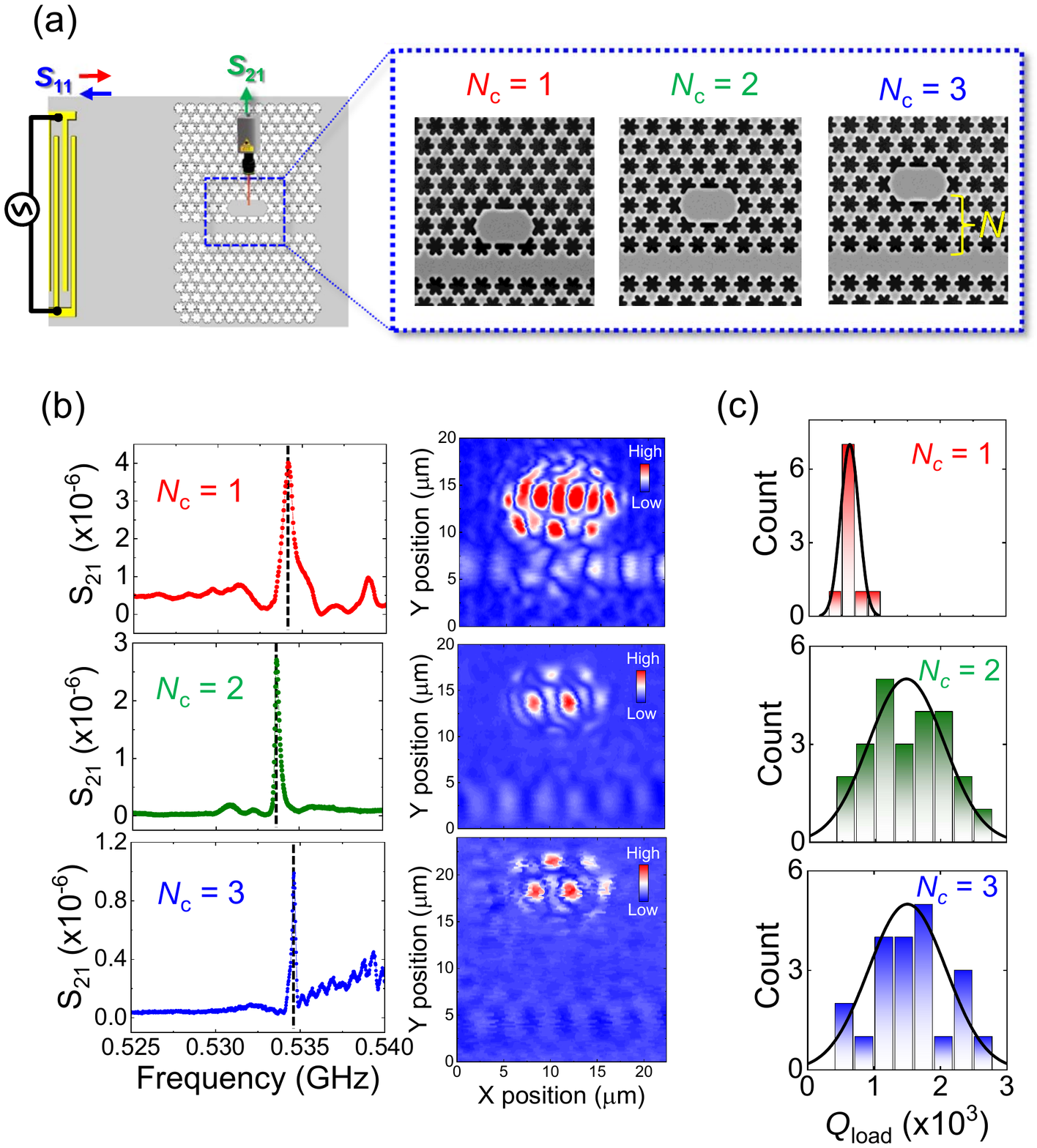}
		\vspace{-3.5cm}
		\caption{\textbf{(a)} The measurement setup of a cavity-waveguide structure where an L2 cavity is side-coupled to a line-defect waveguide with spatial separation between a cavity and a waveguide $N_{\rm c}$ = 1, 2 and 3 and they are shown in the left, middle and right insets respectively. \textbf{(b)} The spectral responses of $S_{21}$ transmissions in the L2 cavities in the configurations $N_{\rm c}$ = 1 (top), 2 (middle) and 3 (bottom). The observed resonance peaks originate from the M1 mode, which are confirmed by amplitude real-space mapping as shown in the right insets. \textbf{(c)} Loaded $Q$ histogram of the L2 cavity system in these configurations.}
		\label{fig 5}
		\vspace{-0.5cm}
	\end{center}
\end{figure}

\newpage

\begin{figure}[t]
	\begin{center}
		\vspace{-0.2cm}\hspace{-2.3cm}
		\includegraphics[scale=0.51]{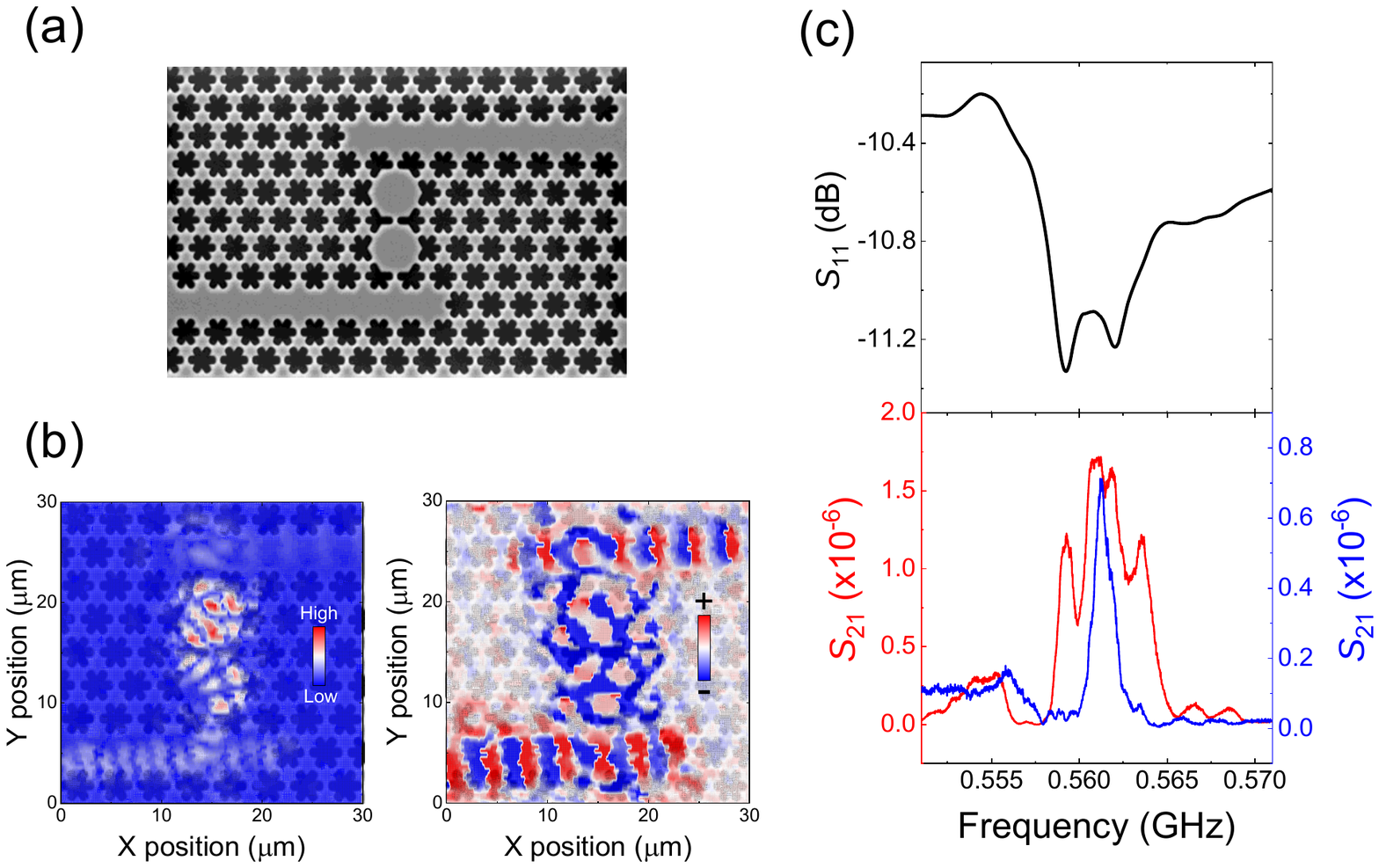}
		\vspace{-8.5cm}
		\caption{\textbf{(a)} An SEM image of a PnC-based microwave filter, where two L1 cavities are sandwiched by waveguides. The bottom and top waveguides are used as input and output waveguides, respectively. \textbf{(b)} The spatial distribution of the amplitude (left) and phase (right) of hypersonic transmission at 0.561 GHz. \textbf{(c)} $S_{11}$ (top) and $S_{21}$ (bottom) spectra measured at an input (red line) and an output (blue line) waveguide.}
		\label{fig 6}
		\vspace{-0.5cm}
	\end{center}
\end{figure}

\end{document}